\documentclass[]{aa}
\usepackage{graphicx}
\usepackage{epstopdf}

\usepackage{txfonts}
\begin{document}
\title{Spitzer Secondary Eclipses of Qatar-1b}
    \author{Emily Garhart\inst{1}\thanks{Present address: School of Earth and Space Exploration, Arizona State University, Tempe, AZ 85281, USA}, Drake Deming\inst{1}, Avi Mandell,\inst{2}  Heather Knutson, \inst{3} and Jonathan~J.~Fortney \inst{4}}
       \institute{Dept. of Astronomy, University of Maryland, College Park, MD 20742, USA\\
                  \email{egarhart@terpmail.umd.edu}
             \and
                 Planetary Systems Laboratory, NASA Goddard Space Flight Center, Greenbelt, MD 20771, USA
                 \and
                 Division of Geological and Planetary Sciences, California Institute of Technology, Pasadena, CA 91125, USA
                 \and
                 Department of Astronomy and Astrophysics, University of California, Santa Cruz, CA 95064, USA
                 }
 
  \abstract
   {}
   {Previous secondary eclipse observations of the hot Jupiter Qatar-1b in the Ks band suggest that it may have an unusually high day side temperature, indicative of minimal heat redistribution.  There have also been indications that the orbit may be slightly eccentric, possibly forced by another planet in the system.  We investigate the day side temperature and orbital eccentricity using secondary eclipse observations with Spitzer.}
   {We observed the secondary eclipse with Spitzer/IRAC in subarray mode, in both 3.6 and 4.5\,$\mu$m wavelengths. We used pixel-level decorrelation to correct for Spitzer's intra-pixel sensitivity variations and thereby obtain accurate eclipse depths and central phases.}
   {Our 3.6\,$\mu$m eclipse depth is 0.149 $\pm$ 0.051\% and the 4.5\,$\mu$m depth is 0.273 $\pm$  0.049\%. Fitting a blackbody planet to our data and two recent Ks band eclipse depths indicates a brightness temperature of 1506\,$\pm$\,71K. Comparison to model atmospheres for the planet indicates that its degree of longitudinal heat redistribution is intermediate between fully uniform and day-side only. The day side temperature of the planet is unlikely to be as high (1885K) as indicated by the ground-based eclipses in the Ks band, unless the planet's emergent spectrum deviates strongly from model atmosphere predictions. The average central phase for our Spitzer eclipses is $0.4984 \pm 0.0017$, yielding $e\cos{\omega} = -0.0028\pm 0.0027$.  Our results are consistent with a circular orbit, and we constrain $e\cos{\omega}$ much more strongly than has been possible with previous observations.}
   {}


\maketitle
\section{Introduction}
    Qatar-1b was the first exoplanet discovered with the Qatar Exoplanet Survey by Alsubai et al. (2011). It orbits a metal-rich K-dwarf star with a period of 1.42 days at an orbital separation of 0.023 AU. Revised estimates of the hot Jupiter's mass and radius by Covino et al. (2013) show that Qatar-1b has a mass of $\sim\,1.33\,M_{Jup}$ and a radius of $\sim\,1.18\,R_{Jup}$.
    Radial velocity observations (Covino et al. 2013), and a secondary eclipse detection with the Calar Alto Observatory in the Ks band (Cruz et al. 2016), allow a slight orbital eccentricity of Qatar-1b. The former found an eccentricity of $ e = 0.020_{-0.010}^{+0.011}$ while the latter obtained $ e\,cos\,\omega$ of $-0.0123_{-0.0067}^{+0.0252}$. 
    
    Previous secondary eclipse observations in the Ks band with the Canada-France-Hawaii Telescope by Croll et al. (2015) did not find any evidence of an eccentric orbit. Alsubai et al. (2011) also favored a circular orbit, although they reported an upper limit of $e = 0.24$. von~Essen et~al. (2013) observed long-term transit timing variations (TTV) over $\sim 190$ days that could indicate the presence of a second body in the Qatar-1 system. This hypothetical perturber could potentially maintain a non-circular orbit of Qatar-1b in the presence of tidal circularization.
    
    However, two other transit analyses by Mislis et al. (2015) and Maciejewski et al. (2015) were inconclusive or had a firm non-detection of TTVs, respectively. Mislis et al. (2015) needed more precise data to detect the long-term TTVs found in von~Essen et al. (2013). Maciejewski et al. (2015) concluded that the Qatar-1 system lacks any tertiary body able to produce periodic transit variations greater than 1 minute. Their analysis of the Qatar-1b orbital and planetary parameters do agree, however, with those of Covino et al. (2013) and von~Essen et al. (2013).  Most recently, Collins et al. (2017) found no evidence for sinusoidal transit timing variations with an upper limit of $\sim\,25$\,seconds, and von~Essen et al. (2017) report transmission spectroscopy of the exoplanetary atmosphere, with evidence for a clear atmosphere.
    
   The secondary eclipse depth of Qatar-1b derived by Cruz et al. (2016) implies a brightness temperature of 1885K.  Hot transiting planets are potential early targets for JWST, increasing the interest in this system.  We here report the secondary eclipse observed with Warm Spitzer in the 3.6 and 4.5\,$\mu$m bands, to make an independent and more precise assessment of the temperature and orbital eccentricity. Sec. 2 discusses the Spitzer observations, and photometry extracted from the data. Sec. 3 describes our data analysis process to obtain eclipse depths using pixel-level decorrelation (PLD, Deming et al. 2015). In Sec. 4 we discuss our interpretations of the temperature and orbital eccentricity of Qatar-1b.
    
\section{Observations}
    Our two secondary eclipses were observed with Spitzer/IRAC in subarray mode under the program 10102 (PI: D. Deming). Each channel has a total of 34,560 exposures of 0.36 seconds, separated into data cubes of 64 frames each. The 3.6\,$\mu$m data were observed on 2014 November 26 and the 4.5\,$\mu$m data on 2014 December 1.  Because this was rated as a low priority program, we facilitated scheduling by specifying wide (40-minute) timing windows.  The 4.5\,$\mu$m observations started near the end of the window, resulting in a minimal pre-eclipse baseline.  The 3.6\,$\mu$m pre-eclipse baseline is longer, but some of those data had to be omitted from the analysis (see below). 
    
    We initially cleaned the data of energetic particle hits and hot pixels using a $\> 4\sigma$ pixel rejection, based on a median filter in time. These pixels were replaced with the median value of the frame. We construct a histogram of the background values outside of a 8x8 pixel mask on the star, and fit a Gaussian to this histogram in order to measure and subtract the sky background. For the purpose of aperture photometry, we then find the center of the star using both a 2D Gaussian fit and a center of light method. 
    
    We perform aperture photometry using IDL Astronomy User Library's \emph{aper} procedure with both fixed and variable aperture radii. The fixed aperture radii are from 1.6 to 3.5 pixels, incremented by 0.2 pixels. The variable aperture radii are computed using the noise-pixel parameter, $\sqrt{\beta}$ (Lewis et al. 2013, Beatty et al. 2014) added to a constant value ranging from 0.0 to 2.0 pixels. We thereby produce four versions of the photometry of each eclipse, using fixed vs. variable aperture sizes and Gaussian vs. center-of-light centering. We omitted the first 45 minutes of the 3.6\,$\mu$m data due to a significant initial ramp in flux, as revealed by multiple binned-data points lying consistently below our first trial fits. The Qatar-1b 4.5\,$\mu$m data need no trimming.

    \begin{figure}[ht!]
        \includegraphics[scale=0.5]{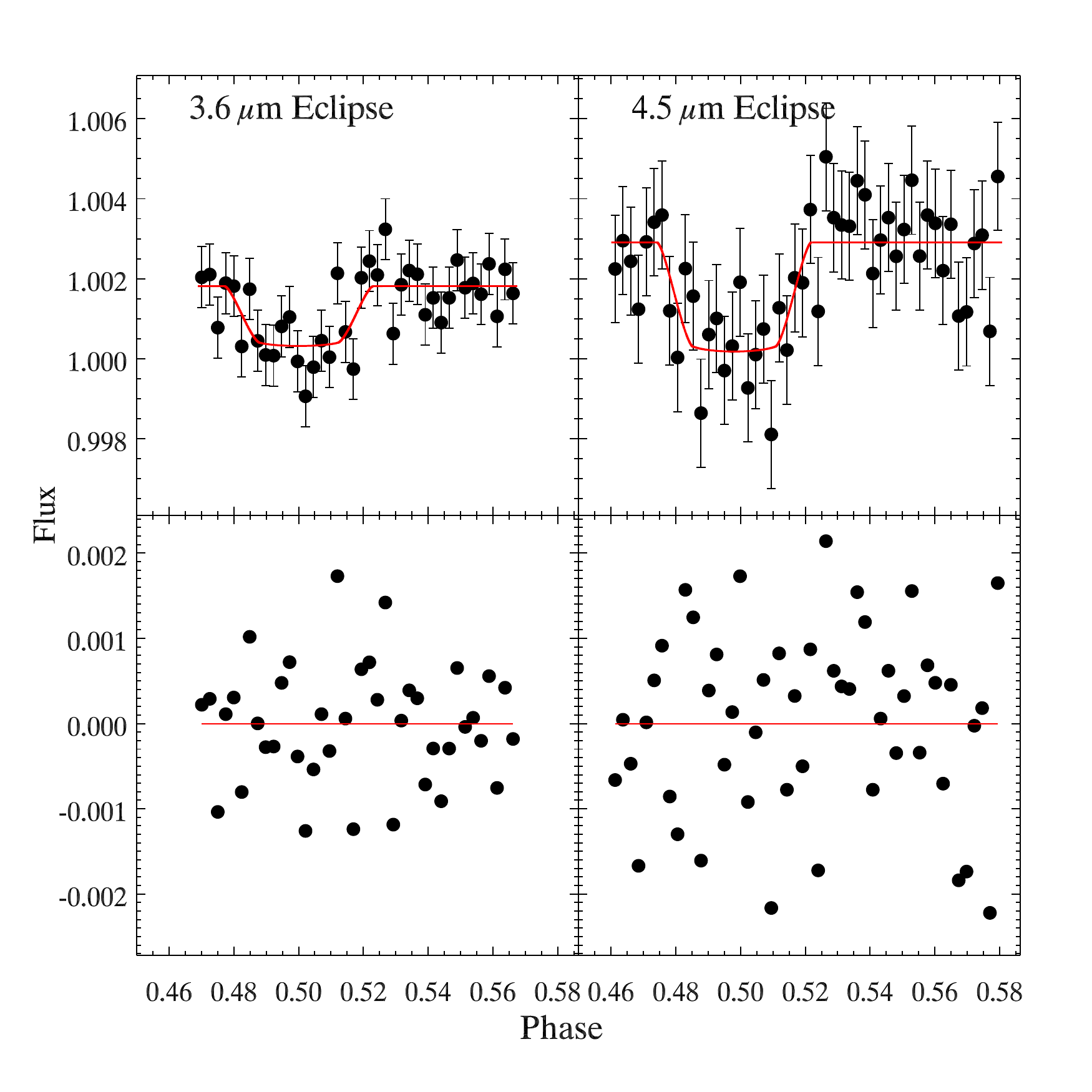}
        \caption{Secondary eclipse binned photometry after PLD analysis in both 3.6 and 4.5\,$\mu$m bands, normalized to unity in eclipse. Best fit solutions, with the intra-pixel effect removed, are overlaid in red. These are binned data, and the error bars are calculated from the scatter in each bin. The bottom panels show the residuals of each fit on an expanded scale.} 
    \end{figure}
    
\section{Data Analysis and Results}
    
    In order to remove the intra-pixel sensitivity fluctuations that strongly affect Spitzer eclipse observations, we use the PLD method described in Deming et al. (2015) (see Dittmann et al. 2017, Kilpatrick et al. 2017, Buhler et al. 2016, Fischer et al. 2016, and Wong et al. 2016 for recent uses of PLD), with two updates.  First, we use 12 pixels encompassing each stellar image as basis vectors in the decorrelations, rather than the 9 pixels used by Deming et al. (2015).  We find that the additional 3 pixels have non-negligible flux levels and they improve the decorrelations.  These 12 pixels form a 4x4 box without corners around the stellar center. Also, the original fit criterion used by Deming et al. was to minimize the $\chi^2$ value in the Allan deviation relation that defines the behavior of the residuals from the fit as a function of bin size (see Sec.~3.3 of Deming et al. 2015). We have modified that fit criterion to seek the minimum raw scatter in the Allan deviation relation, rather than the minimum $\chi^2$.  This slight update in the best-fit criterion has the effect of placing more weight on the longer time scales present in the data, specifically on times comparable to the duration of the eclipse.  
    
    \subsection{Modeling the Secondary Eclipse}
    
    We perform the analysis described here for each of the four sets of photometry. Each set of photometry contains multiple aperture sizes.  We initially perform a multivariate linear regression on the unbinned data using the median value of aperture size.  This regression loops over phase to locate the eclipse and produce a preliminary estimate of its central phase, minimizing the $\chi^2$ of the fit to the unbinned data.  We calculate phase from the observed barycentric time using the ephemeris of Collins et al. (2017).  Subsequent regressions hold the central phase constant, and use multiple combinations of binning and aperture size to find the combination that produces the minimum $\chi^2$ in the fit to the binned data. The data are fit with a function described by:
    
    \begin{equation}
        \Delta S^{t} = \sum_{i=1}^{N} c_{i}\hat{P}^{t}_{i} + DE(t) + ft + h
    \end{equation}
    
    $\Delta S^{t}$ is the total fluctuation in the brightness of the star at time $t$ from all sources. The pixel intensities, $\hat{P}^{t}_{i}$ are normalized so they are independent of the eclipse. $DE(t)$ is the eclipse depth times the eclipse shape and $h$ is a constant offset. We compute the eclipse shape using the Mandel \& Agol (2002) procedure, and we explored using Gaussian priors on the orbital inclination and $a/R_s$ parameters, as well as fixing those parameters at the values given by Alsubai et al. (2011).  The linear term, $ft$, fits the temporal instrumental baseline for both 3.6 and 4.5\,$\mu$m, and we excluded a quadratic term in time based on a Bayesian Information Criterion (Schwarz 1978).  The regressions find the best fit of Eq.(1) for a given aperture and bin size combination.  The fitting code then selects the best aperture radius and bin size combination by minimizing the scatter in the Allan deviation relation (Allan 1966), constraining the slope to -0.5 (i.e., residuals whose scatter decreases as the inverse square root of bin size).  This yields a solution that considers all of the time scales represented in the data, as discussed in Sec.~3.3 of Deming et al. (2015). For these eclipses, our best fits used binning over 336 and 544 points, and photometry apertures having constant radii of 1.6 and 2.0 arc-sec,  at 3.6- and 4.5\,$\mu$m respectively.
    
    We repeated the process described above for each of the four sets of photometry (two centroiding methods, each using fixed versus variable aperture radii). We adopt our final result based on which of the four required the smallest re-scaling ratio of the best-fit photometric scatter to the photon noise (see below). The fixed aperture radii resulted in lowest errors for both of our secondary eclipses. The 3.6\,$\mu$m data needed a smaller re-scaling of the photometric error using the center-of-light method and priors, while the 4.5\,$\mu$m eclipse had lower errors with Gaussian centroiding and fixed orbital parameters.  The 3.6\,$\mu$m re-scaling factor was 1.30, versus 0.99 at 4.5\,$\mu$m (errors 30\% greater, and closely equal to the photon noise, respectively).  We verified that other versions of the photometry and other fitting procedures (e.g., prioring the orbital parameters, or not) did not produce eclipse depths that disagree significantly with the results reported here.

    \begin{figure}[ht!]
        \includegraphics[scale=0.5]{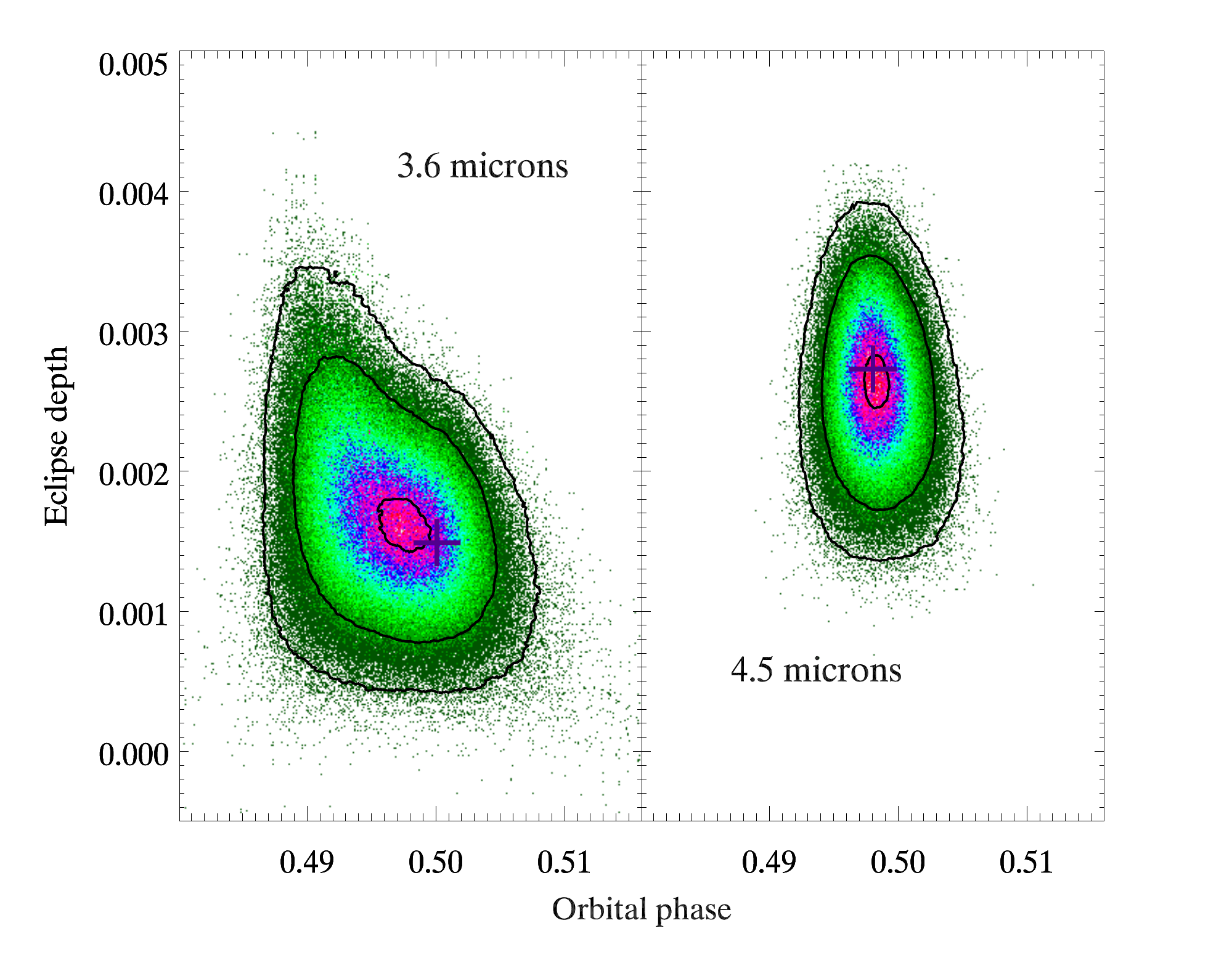}
        \caption{Joint posterior distributions of the central phase and eclipse depth from the 3.6\,$\mu$m MCMC analysis (left panel), and 4.5\,$\mu$m (right panel), plotting each point from the MCMC and indicating the density of points by color. The contours (outward to inward) encompass 99\%, 10\%, and 1\% of the points.  The crosses mark the best-fit solutions determined by minimal scatter in the Allan deviation relation (see text).}
    \end{figure}
    
    \begin{figure}[ht!]
        \includegraphics[scale=0.45]{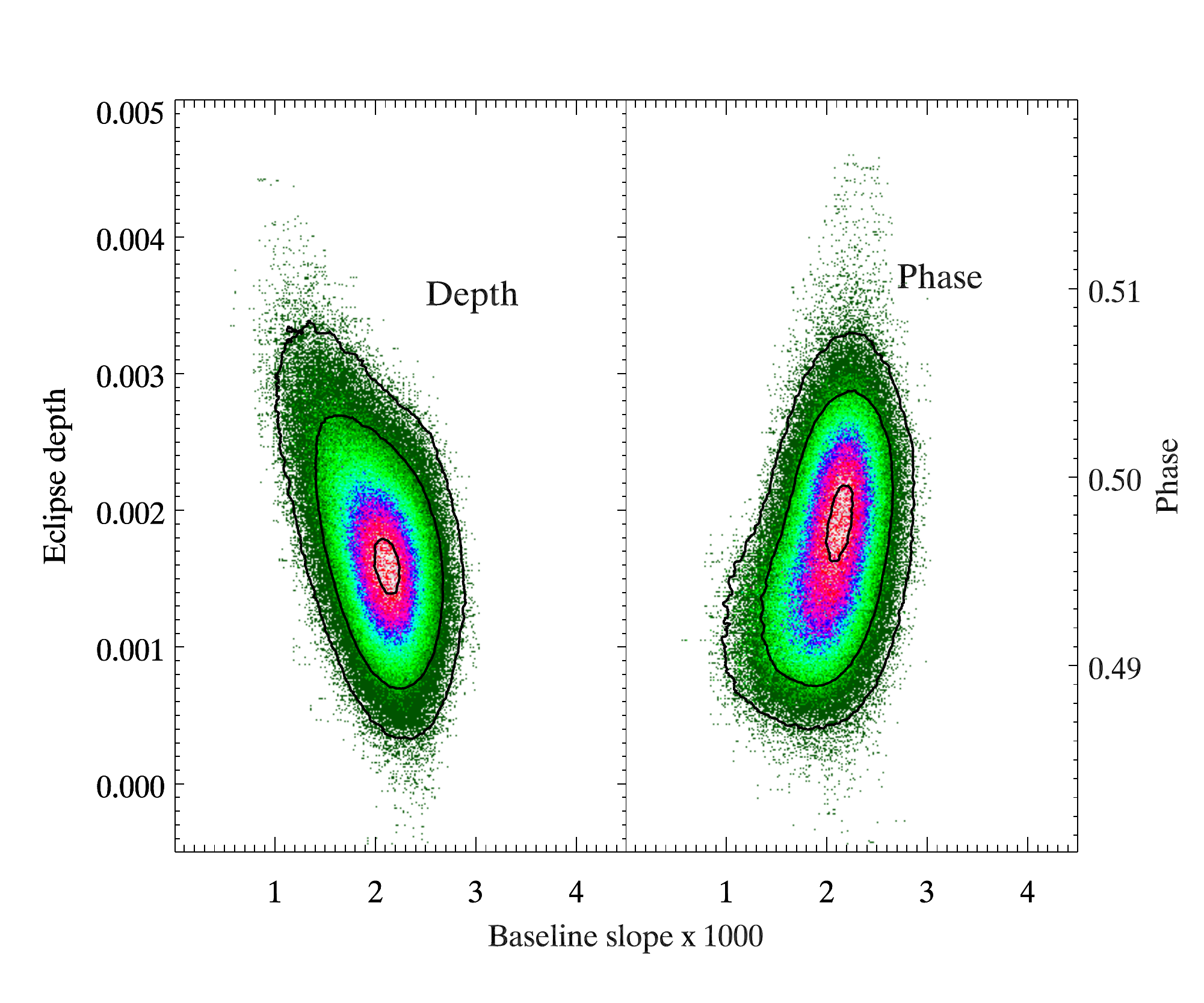}
        \caption{Joint posterior distributions of the 3.6\,$\mu$m eclipse depth and central phase, versus the slope of the baseline ($f$ in Equation 1, in units of hours$^{-1}$). }
    \end{figure}

    We used a Markov Chain Monte Carlo (MCMC) procedure (Ford, 2005) to estimate the errors on the central phase, and eclipse depth in both channels.  The MCMC was split into three main components to reduce the computation time. We first run a burn-in period of 10,000 steps to adjust the step sizes for each fitted parameter.  Also, we re-scale the photometric errors so that the reduced $\chi ^{2}$ of the fit to the binned data is close to unity. We then run the bulk of the MCMC analysis on the binned data for 800,000 steps in order to sample the entire parameter space. The MCMC is sometimes able to find a better fit than did the regressions. The regressions that choose the best bin size and aperture radius do not vary the central phase of the eclipse like the MCMC procedure does. Therefore it is possible for the MCMC to slightly improve the fit, by dithering the eclipse phase and depth simultaneously.

   The secondary eclipses in the 3.6 and 4.5\,$\mu$m bands are shown in Figure 1. The two frames show the binned eclipse data with the PLD fit overplotted.  In Figure 2 we show the joint posterior distributions for central phase versus eclipse depth, at 3.6 and 4.5\,$\mu$m, with crosses marking the best fit values found in the MCMC. Figure~3 shows the posterior distributions for depth and phase at 3.6\,$\mu$m versus the slope of the temporal ramp, in order to show whether the trimming of initial data introduced degeneracies between the eclipse parameters and the ramp.  The PLD eclipse depths, and central phases and times of the eclipses are listed in Table 1.  
   
   In addition to the PLD fits, we also explored simpler fitting procedures, as a check on our results.  The first decorrelations of the intra-pixel effect in Spitzer data (e.g., Charbonneau et al. 2005) used polynomial functions of the $X$ and $Y$ positions of the image centroid as basis vectors.  We implemented polynomial (quadratic) decorrelation fits to both the 3.6- and 4.5\,$\mu$m eclipses, by substituting the $X$, $X^2$, $Y$, and $Y^2$ positions of the image for the pixel coefficients in Eq.(1), with $N=4$, and fitting to photometry that is binned over 32 frames. Unlike the PLD methodology, exploration of different data binning has not been commonly implemented when decorrelating with polynomials.  We therefore made the conservative choice to bin over 32 frames. That divides the original data cubes exactly in two, allowing us to check the internal consistency of the data cubes, while still implementing the advantages of data binning (see Sec.~3 of Deming et al. 2015).  However, we checked many other choices of bin size and verified that the results for the polynomial fits are not sensitive to the bin size used in the decorrelation process.

   \begin{figure}[ht!]
        \includegraphics[scale=0.45]{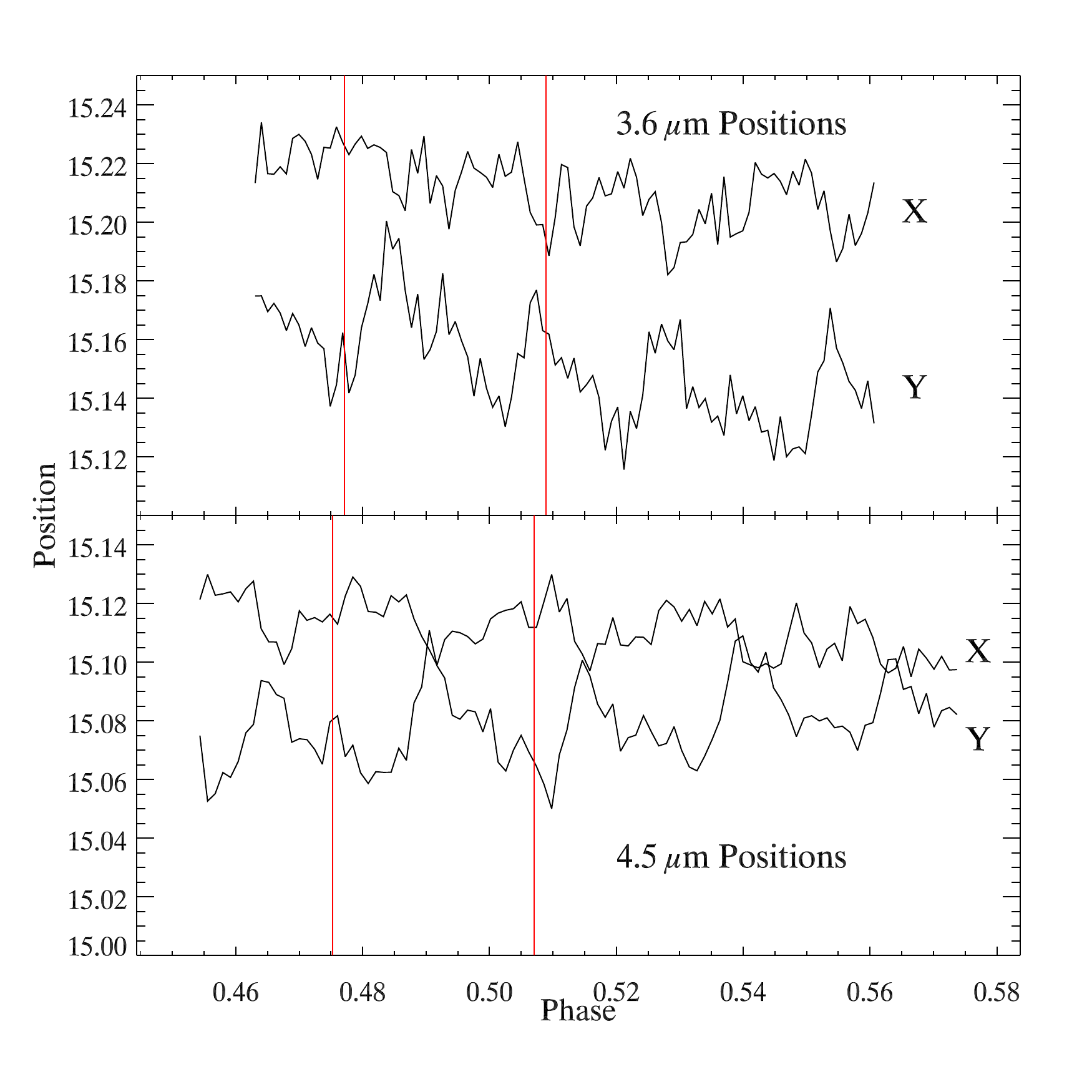}
        \caption{$X$ and $Y$ positions of the image versus orbital phase for both the 3.6- and 4.5\,$\mu$m eclipses.  The vertical red lines are the phases of ingress and egress. }
    \end{figure}

     The polynomial fits are poorer than the PLD solutions. Rebinning the residuals for the polynomial solutions to the same bin sizes selected by our PLD code, we find ratios of the scatter to the photon noise of 1.82 and 1.18, at 3.6- and 4.5\,$\mu$m respectively, for the polynomial solutions, versus 1.30 and 0.99 for the PLD results. In spite of the larger error, the 3.6\,$\mu$m polynomial eclipse depth is consistent with the PLD result, but the 4.5\,$\mu$m result differs by $2\sigma$ from the PLD fit. (The results and differences are discussed further in Sec.~3.2).  Although $2\sigma$ would not be a sufficient difference to prove a discrepancy, we were motivated to investigate the 4.5\,$\mu$m fits in more depth.  
   
     For these eclipses the image motion is small, less than 0.1-pixels total motion over the full duration of both eclipses, as shown in Figure~4.  Although the 3.6\,$\mu$m eclipse shows the (normal) correlation between the stellar flux and the image position (especially in the $Y$-coordinate), we found an unusually weak correlation between flux and image position at 4.5\,$\mu$m.  Specifically, the Pearson correlation coefficients between position and 4.5\,$\mu$m flux are -0.118 and 0.022 (for $X$ and $Y$, respectively).  Although those values are arguably statistically significant given the large number of data points, the correlations are much weaker than we usually see in Spitzer data at 4.5\,$\mu$m.  Therefore we explored a third method to derive the eclipse depth at 4.5\,$\mu$m: we simply bin the photometry and fit an eclipse with an exponential temporal ramp, using no decorrelation of any kind.  The interesting results of this 'simple fit' are illustrated and discussed in Sec. 3.2.

    \begin{figure}[ht!]
        \includegraphics[scale=0.5]{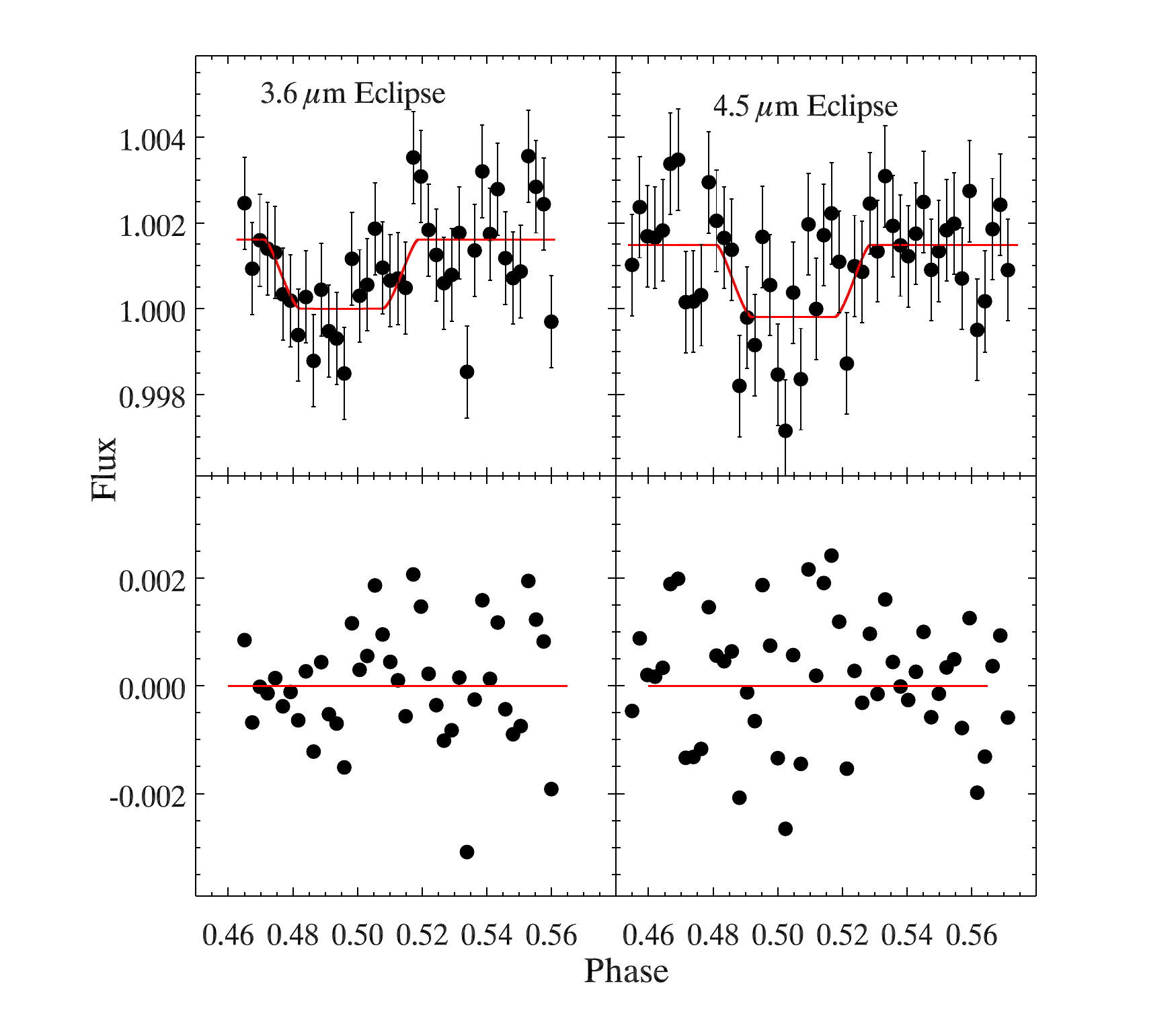}
        \caption{Eclipses derived by removing the intra-pixel variation in the photometry via decorrelating versus the $X$, $X^2$, $Y$, and $Y^2$ positions of the image. These fits are poorer than the PLD fits shown in Figure~1.}
    \end{figure}

    \subsection{Results}
    
     The best fit values for eclipse depth and central phase in the PLD solutions are close to the peak of the posterior distributions {for both eclipses (Figure~2) . (None of the conclusions of this paper would change if we adopted the peak of the joint posterior distributions for central phase and eclipse depth, rather than the best-fit values chosen by our code.)  Inspecting the MCMC results, we found no correlations between any of the pixel coefficients and the eclipse depths in either band.  We do find weak correlations between the time coefficient ($f$) in Eq.(1), and the 3.6\,$\mu$m eclipse depth and phase (Pearson correlation coefficients of -0.48 and +0.41 respectively, shown on Figure~3).  The MCMC error on the eclipse depth includes these correlations. The correlations arise from the scarcity of pre-eclipse baseline (see Figure~1), and no such correlations occur at 4.5\,$\mu$m.

    The polynomial eclipse fits are shown in Figure~5, and tabulated in Table~2.  The visual appearance of the polynomial fits are noticeably worse than the PLD solution (compare Figures 1 and 5).  Given the $2\sigma$ difference between the PLD and polynomial solutions at 4.5\,$\mu$m, we turn to the results from the simple fit, illustrated in Figure~6 and also included in Table~2.  In this case, we used an exponential temporal ramp because the data are sharply increasing before ingress and essentially flat after egress.  We verified that the exponential is superior to a quadratic or linear ramp based on the Bayesian Information Criterion.  We fit to the binned data illustrated on Figure~6, and we calculated the error in eclipse depth and central phase using a bootstrap Monte Carlo procedure.  The results of this simple fit support the PLD solution (Table~1) as opposed to the polynomial result at 4.5\,$\mu$m (Table~2).  It is interesting that the PLD solution did not require the exponential ramp: when we implement that ramp in the PLD code the exponential parameters collapse to produce very close to a linear ramp.  We point out that an exponential ramp that is caused by only a subset of the pixels would tend to be removed by the pixel coefficients in Eq.(1), and would not necessarily propagate as a purely temporal effect.

    \begin{figure}[ht!]
        \includegraphics[scale=0.5]{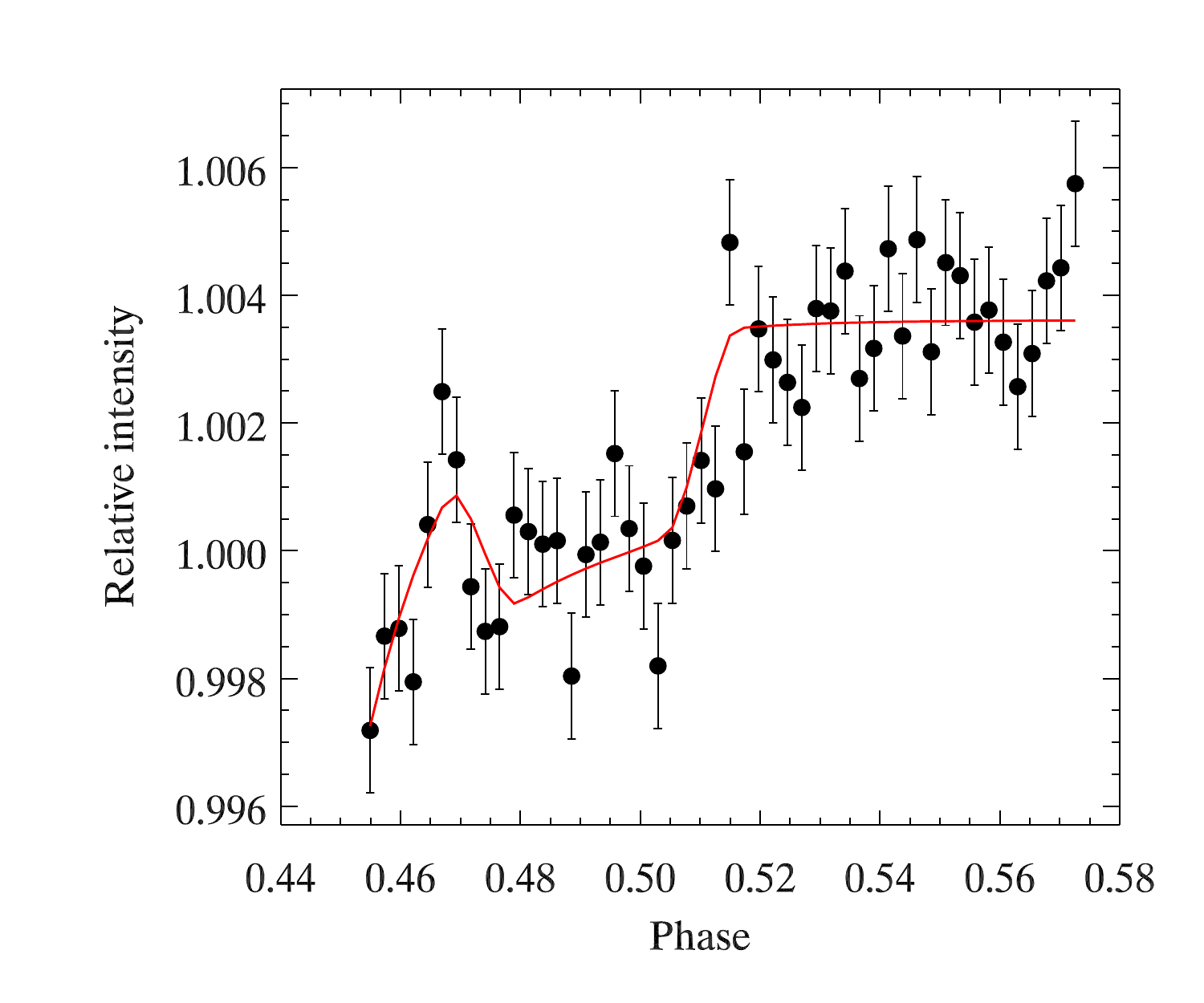}
        \caption{Result of a simple fit at 4.5\,$\mu$m, where an eclipse curve of variable depth and central phase is fit to these binned photometry points, including an exponential ramp, but no attempt to decorrelate any intra-pixel effects. }
    \end{figure}

    The results of the simple fit do not depend on any specific model of the intra-pixel effect, because the simple fit does not use an intra-pixel correction.  Because the simple model-independent fit agrees with the PLD solution, we conclude that the PLD fit is more reliable than the polynomial result at 4.5\,$\mu$m.  Further reasons to reject the polynomial solution at 4.5\,$\mu$m are the weak correlation of flux with image motion, and the larger ratio of scatter to photon noise as compared to the PLD value (PLD reaches the photon limit at 4.5\,$\mu$m).  We therefore adopt the Table~1 eclipse depths and central phases as representing Qatar-1b, and we now discuss the implications of those values.

    \begin{table}[]
    \caption{Qatar-1b eclipse depths and central phase and time in both 3.6 and 4.5\,$\mu$m. These are the PLD values, and we adopt them as our results.}
    \tabcolsep=0.11cm
    \begin{tabular}{lll}
    Wavelength  & 3.6 $\mu$m & 4.5 $\mu$m \\
    \hline \hline
    Eclipse Depth (\%) & 0.149 $\pm$ 0.051 & 0.273 $\pm$  0.049\\
    Central Phase & 0.50009 $\pm$ 0.0041 & 0.49805 $\pm$  0.0019\\
    BJD(TDB) & 56987.4262 $\pm$ 0.0165 & 56993.1034 $\pm$ 0.0158\\
    \end{tabular}
    \end{table}

    \begin{table}[]
    
    \caption{ Qatar-1b eclipse depths and central phase based on polynomial decorrelations, and the simple fitting procedure at 4.5\,$\mu$m (see text).  These values are not adopted as our results; they are used as a check on the PLD results given in Table~1.}
    \tabcolsep=0.11cm
    \begin{tabular}{lll}
    Wavelength  & 3.6 $\mu$m & 4.5 $\mu$m \\
    \hline \hline
    Poly Eclipse Depth (\%) & 0.161 $\pm$ 0.040  &  0.168 $\pm$ 0.051  \\
    Poly Central Phase & 0.4940 $\pm$ 0.0034  &  0.4989 $\pm$ 0.0069   \\
    Simple Eclipse Depth (\%) & --- &  0.326 $\pm$ 0.047  \\
    Simple Central Phase & --- &  0.4918 $\pm$ 0.0014 \\
    \end{tabular}
    
    \end{table}

    \begin{figure}[ht!]
        \includegraphics[scale=0.5]{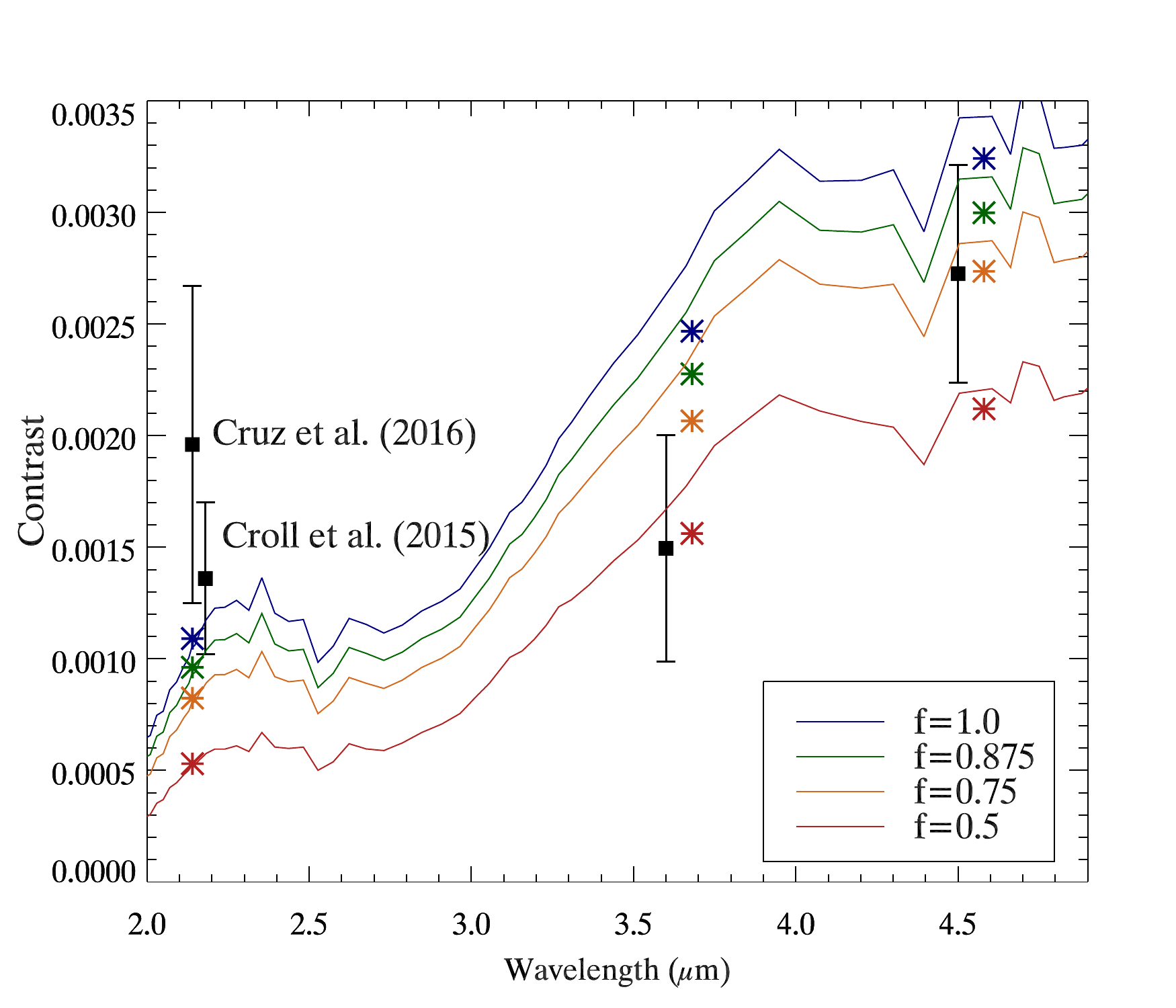}
        \caption{Fit to our observed eclipse depths and two Ks band depths found by Croll et al. (2015) and Cruz et al. (2016), adopting model atmospheres for the planet (Fortney et al. 2006) and an ATLAS model atmosphere for the star.   The four planetary models correspond to different heat redistribution coefficients, with f=1.0 being only day-side redistribution and f=0.5 being planet wide; the models contain no TiO/VO opacity and have no temperature inversions.  Contrast (ordinate) is the planetary flux divided by the stellar flux. The Croll et al. result is offset in wavelength from 2.14\,$\mu$m for clarity. The asterisks, also offset from the two Spitzer channels, are the contrasts that result from integrating the stellar and planetary fluxes over the Ks and Spitzer bandpass functions.} 
    \end{figure}
    
\section{Discussion}
    \subsection{Day-side Temperature}
    We fit a blackbody planet of varying day side temperature to our 3.6 and 4.5 $\mu m$ eclipse depths and the Ks band eclipse depths reported in Cruz et al. (2016) and Croll et al. (2015).  We also compare to non-gray model atmospheres for the planet (Figure~4) with different heat re-distribution efficiencies (Fortney et al. 2006).  In all cases, we represented the star using an ATLAS model\footnote{http://kurucz.harvard.edu/grids.html} with an effective temperature of 5000K and log $g$ of 4.5. Covino et al. (2013) reported a stellar effective temperature of 4910K, so we adjust the stellar model flux to that temperature.
    
    Varying the blackbody temperature, we find the brightness temperature, $T = 1506K \pm 71K,$ for the planet is the best fit to the 4 observed eclipse depths. This value agrees with the $T_{eq}$ estimated in both the Alsubai et al. discovery paper as well as Covino et al. (2013) within the errors. The latter used a uniform heat distribution, $\varepsilon = 1$, and a zero Bond albedo, $A_B = 0$, to estimate a planetary equilibrium temperature of 1389K using (Cowan \& Agol 2011):
    
    \begin{equation}
        T_{eq} = T_s(\dfrac{R_s}{a})^{1/2}(1-A_B)^{1/4}(\dfrac{2}{3}-\dfrac{5}{12}\varepsilon)^{1/4}
    \end{equation}
    
    We calculate the maximum equilibrium temperature, with no heat redistribution $T_{\varepsilon = 0}$, as $1775K \pm 39$K. In Cowan \& Agol (2011) Figure 7, the maximum day-side temperature is plotted against the ratio, $T_d/T_0$, of the observed equilibrium temperature to the temperature at the sub-stellar point. That relation is diagnostic for the degree of longitudinal heat redistribution on the planet. Assuming a circular orbit, $T_0 = T_{eff} /\sqrt{a/R_s} = 1966K \pm 41K$ using a/$R_s$ from Collins et al. (2017). With our blackbody fit temperature, $T_d/T_0$ is 0.766 $\pm 0.039$. This places Qatar-1b with the majority of planets between zero heat re-circulation and a uniform planet.  
    
    We now discuss the planetary model spectra shown on Figure~7.  In this case we compare the range of models to the observations, but we do not explicitly vary the models to attempt a fit.  The models partially account for the apparent discrepancy between the Ks band and Spitzer eclipse depths: they predict an enhanced eclipse depth near 2.1\,$\mu$m compared to a blackbody, because of a minimum in the opacity at that wavelength.  While no model accounts for all of the observations, the $f=0.75$ model is the best of the four. The $\chi^2$ between that model and all of the observations is 8.72 for 4 degrees of freedom, disfavoring the model at only at the 93\% confidence level (usually much higher confidence is required for rejection).  We conclude that the $f=0.75$ model accounts for the observations to a (minimally) acceptable degree, and that the degree of heat circulation on Qatar-1b is similar to most hot Jupiters.  We also conclude that the day side temperature of the planet is unlikely to be as high as indicated by the ground-based eclipses in the Ks band. This planet is a favorable target for JWST observations because the ground- versus space-borne results suggest unusual modulation in the spectrum, not because it is strongly heated.
        
    \subsection{Secondary Eclipse Timing}
    We expect the observed center of eclipse to occur at phase 0.5002 because of the 23.4 second light travel time across the orbit. The offset of the observed central phase from this value can indicate an eccentricity in the Qatar-1b orbit. We calculate the $e \cos{\omega}$ value with
    \begin{equation}
         e\cos{\omega}  = \pi ( \dfrac{\Delta \phi}{1+\csc^2(i)}),
    \end{equation}
    
    (Wallenquist 1950, López-Morales et al. 2010). $\Delta \phi$ is the difference between the observed central phase and the value of 0.5002 for a circular orbit. For each wavelength we find $\Delta \phi_{3.6} = -0.0001 \pm 0.0041$ and $\Delta \phi_{4.5} = -0.0022 \pm 0.0019$ where the errors are the standard deviation of the central phase posterior distributions. The ephemeris uncertainty does not add significant error to the eclipse phase. The average eclipse phase for the two wavelength bands, weighting each by the inverse of its variance, is $\Delta \phi = -0.0018 \pm 0.0017$, and  $e\cos{\omega} = -0.0028 \pm 0.0027$.  Our results are consistent with a circular orbit, with the Spitzer results enabling the limit on $e\cos{\omega}$ to be improved by about a factor of 6 over the results from Cruz et al. (2016).
    
\section{Conclusions}

    Our measured secondary eclipse depths in the Spitzer bands at 3.6 and 4.5\,$\mu$m indicate that the Qatar-1b's day side temperature is 1506\,$\pm$\,71K, not as hot as suggested by ground-based observations in the Ks band.  The planet is nevertheless an attractive target for JWST spectroscopy because the secondary eclipse photometry (i.e., Ks band versus Spitzer) suggests significant modulation in the day side emergent spectrum. Comparisons with model atmospheres indicate that the planet re-distributes heat to a degree intermediate between uniform and day side only. Timing of Spitzer's secondary eclipses are consistent with a circular orbit, with our limit on $e\cos{\omega} = -0.0028\pm 0.0027$ being about 6 times more stringent than previous results.

\end{document}